# Inverse Structural Design of Graphene/Boron Nitride Hybrids by Regressional GAN

Yuan Dong[1#], Dawei Li[1#], Chi Zhang[1], Chuhan Wu,[2] Hong Wang[1], Ming Xin[1], Jianlin Cheng[2], and Jian Lin[1,2,3*]

[1]Department of Mechanical and Aerospace Engineering
[2]Department of Electrical Engineering and Computer Science
[3]Department of Physics and Astronomy
University of Missouri, Columbia, Missouri 65211, USA
[*]E-mail: LinJian@missouri.edu (J. L.)
[#]Authors contributed equally to this work.

**Abstract**

Inverse design of materials with desired properties is currently laborious and heavily relies on intuition of researchers through a trial-and-error process. The massive combinational spaces due to the constituent elements and their structural configurations are too overwhelming to be all searched even by high-throughput computations. Herein, we demonstrated a novel regressional generative adversarial network (RGAN) for inverse design of representative two-dimensional materials, graphene and boron-nitride (BN) hybrids. A significant novelty of the proposed RGAN is that it combines the supervised and regressional convolutional neural network (CNN) with the traditional unsupervised GAN, thus overcoming the common technical barrier in the traditional GANs, which cannot generate data associated with given continuous quantitative labels. The proposed RGAN enables to autonomously generate graphene/BN hybrids with any given bandgaps. Moreover, the generated structures exhibit high fidelity, yielding bandgaps within ~ 10% $MAE_F$ of the desired bandgaps as cross-validated by density functional theory (DFT) calculations. Further analysis by principle component analysis (PCA) and modified locally linear embedding (MLLE) methods on the latent features encoded by the regressor reveals that the generator has successfully generated structures that followed the statistical distribution of the real structures. It implies the possibility of the RGAN in recognizing physical rules hidden in the high-dimensional data. This new inverse design methodology would speed up the discovery and development of other 2D materials and beyond.

**Introduction**

One of the core objectives in material research is to correlate material structures with corresponding properties for a goal of designing novel materials with desired performance. However, the searching spaces due to combination of the constituent elements and their structural configurations in the materials are massive, leading to an overwhelming number of potential candidates.[1] Although high-throughput computations,[2] especially those based on the density functional theory (DFT), have enabled the calculation of the structures and properties for various types of materials,[3-6] the design principle is still limited by the search strategy that is based on the intuitions of the material scientists to explore the chemical space in a fixed library. In addition, the computational cost exponentially increases with the size of the systems. As a majority of the functional materials, possess the properties that cannot be simply determined by a few atoms, to accurately predict their properties by the physics-informed models a vast number of atoms and their possible spatial configurations should be considered in the calculations, which would make the calculations either impossible or not cost effective.

Therefore, development of surrogate models, such as machine learning models, has emerged in the era of the big material data for mining the established databases,[7] previous literature[8] and experiments,[9] and new high-throughput experiments.[10] They can be combined with the high-throughput computations for empowering the *in silico* design of materials with a much faster speed.[1] These techniques are expected to suitably tackle complex material challenges arising from the massive combinational chemical spaces or the involved nonlinear processes, thereby widely applied to develop functional materials such as inorganics,[11,12] organic molecules,[13] synthetic polymers,[14] and natural products.[15] Feeding the deep neural networks with the computational data is particularly promising for handling and accelerating material discovery.[16,17] For instance, we recently developed deep convolutional neural network (CNNs) that could accurately predict the bandgaps of configurationally hybridized graphene and boron nitride (BN) with arbitrary BN dopant topographies.[18] The significance of the work was that it solved a challenge of predicting the bandgaps of the hybridized structures with a vast number of possibilities. Because the bandgaps can be continuously evolved based on the topographical configurations of C, B, N atoms in the structures, while the number of possibilities far exceeds what can be exhaustively computed due to the extremely high computational cost of the DFT.

Nevertheless, this work as well as most of other reported works are based on traditional forward design principles, in which the target structures are hypothesized, evaluated against the target properties, and then iteratively refined based upon the results of repeated evaluations. Because this iterative process is trying to search for the optimum structures among the high multivariate and multidimensional parameters, it is hard to reason that the finalized results meet such a goal. In contrast, an inverse design process starts from explicit calculation of the target performance metrics, thus realization of the target structures is much more efficient with higher fidelity.[19] Recently developed generative models, such as variational autoencoders (VAEs) and generative adversarial networks (GANs), have shown great potentials in automating this process.[20] For instance, a VAE was developed to generate molecules.[21] A critical novelty of this work is to connect the multilayer perceptron to the latent continuous vectors that represent the molecules for predicting the chemical properties. Although this reported VAE allows for search of new molecules with certain properties, the network still lacks a mechanism that could autonomously generate structures with predefined or given properties.

GANs possess some characteristics that may have superior advantages to the VAEs in that they allow for interpolation using the vectors through input matrices provided to the generator.

For instance, the GAN was introduced for the inverse design of metasurfaces with desired optical properties.[22] The discriminator and generator were trained simultaneously for the generator to output the similar structures as the ones shown in the specific training classes with the given optical spectra. The inverse design upon the desired spectra was attempted, but a large deviation between the desired spectra and obtained ones was observed. A very recent study by Gupta and Zou used Wasserstein GAN (WGAN) to generate DNA sequences with certain antimicrobial properties or helix structures.[23] A function analyzer (deep recurrent neural network) was adopted to score the antimicrobial properties of the genes streamed from the generator. To score the helix structure properties, a black-box PSIPRED analyzer was used. Those genes with higher scores rated by the analyzer were fed back to train the discriminator. Then the GAN was updated for the next training epoch. The architecture was called a feedback GAN. Having the same limitation as shown in inverse design of metasurfaces,[22] the network cannot generate structures with the input continuous labels. Because the genes are classified into antimicrobial (yes or no) or helix (yes or no). The task of the generator is to obtain new structures similar to those in the training classes. These material design trials involving GAN are still in the region of generating materials within similar and specific classes.

Despite these seminal works, the GAN enabled inverse design of materials with explicitly given properties (represented by continuous labels) in an autonomous manner has been largely underexplored. The biggest challenge of achieving this goal is that the GAN should meet both requirements: 1) it generates novel structures that should be distinguishable from the structures used in the training datasets; 2) it should be able to accurately perform this task based on the input continuous labels, which will require the network to have both conditional and regressional functions. However, the ordinary GAN architectures including conditional GANs[24-26] have limitations in generating data conditioned on continuous quantitative labels, which is exactly how a lot of materials properties, such as bandgaps of the graphene/BN hybrids, look like. Although some regressional GANs achieved initial success,[27] the discriminator is not applied to tell the authenticity of data, and thus requiring extra ad-hoc parameter tuning. Full autonomy of generating the data with high fidelity has yet been achieved.

Given all the limitations addressed in the previously developed GAN structures, we propose a novel GAN architecture that can perform both conditional and regressional tasks so that it can generate material structures with given properties. It is named as regressional GAN (RGAN). The key novelty of this RGAN architecture is that it includes an extra regressional convolutional neural network (named as the regressor) that outputs predicted bandgaps as well as a vector of latent features from a generated or sampled real material structure (Fig. 1). Instead of directly using the generated structure-bandgap pair as the input datasets for training the discriminator, the two sets of the latent features from the generated structure and sampled real material structure are concatenated with their corresponding bandgaps for forming two groups of vectors as the input datasets for training the discriminator. This novel strategy enables better utilization of the automatic authentication from the discriminator. Thus, not only the original functions of GAN are preserved, but also the generated structures can be associated with the required properties, leading to successful training of the RGAN for the inverse design of materials.

To demonstrate its capability in inverse structural design of materials, we present a case study using two-dimensional graphene alloyed with the BN pairs in arbitrary topographies. Our past study shows that these 2D material structures have continuous quantitative bandgaps which are not only determined by the concentration of BN but also by the BN topographies.[18] The trained RGAN can well reflect this structure-property relationship by generating multiple yet distinguished graphene/BN structures upon given bandgap values, showing bandgaps within ~ 10% $MAE_F$ of the desired bandgaps as cross-validated by the density functional theory (DFT) calculations. Further analysis by PCA and MLLE methods revealed the working mechanism of the RGAN network. We envision that the autonomous inverse design of the materials with target properties achieved by the demonstrated RGAN would push a step forward in realizing the self-driven laboratory.[28] Moreover, it infers that the artificial intelligence might enable to catch the physical rules hidden in the data in the high-dimensional space, which is beyond the human's capability. Thus, it would help the scientists to discover new physical and chemical rules to advance material research.

**Design and Training of Regressional GAN**

**Background**. The original architecture of GAN proposed by Goodfellow et al.[29] consists of two competing networks, the generator and the discriminator. The generator is able to synthesize data from a given random noise (Z). The data is then examined by the discriminator to see if it is synthesized (fake) or sampled from the training dataset (real). After the network is trained, a balance between the generator and discriminator is reached and the generated fake data is so real that the discriminator cannot distinguish whether the incoming data is generated from the generator or sampled from the training dataset. The original GAN is unsupervised. To enable on-demand data generation, the labeled information was added as the input condition to the networks. In the conditional GAN (CGAN) for image data generation,[24] the labels are concatenated to the input random noise fed into the generator or concatenated to different color channels (R, G, or B) of images fed to the discriminator. Although CGAN generates data upon the labels, the function is limited to image generation. Moreover, it only suits the labels that are discrete and represented by a qualitative metric. The auxiliary classifier GAN (ACGAN)[26] doesn't append the labels directly to the data fed to the discriminator. Instead, the discriminator is trained to output the authenticity of the fed data as well as predicted labels corresponding to the fed data. Then the predicted labels can be adopted as a part of the objective function for training the ACGAN. However, the ACGAN is still only optimal for problems that can be represented by the discrete and qualitative labels.

Methods like binning, discretizing continuous labels based on its magnitude, or grouping the labels were previously adopted in our practices to generate material structures upon given bandgaps when we tested the ACGAN, but they only resulted in the inaccurate labels. The reason could be that binning labels to the same category without knowing if the materials are sharing similar characteristics would violate the physical laws. A very recent semi-supervised reg-GAN was developed for generating images from the quantitative continuous labels.[27] However, the reg-GAN distinguishes the fake images from the real images by predicting the label first, then compares difference between the predicted label and desired label with a preset ad-hoc range of the number for indicating the real ones. This proposed strategy not only requires presetting the range, thereby limiting broader applications, but also deprives the automatic authentication function afforded by the discriminator. In summary, the general GAN architectures are notorious for its difficulty of training, and faces many problems like model divergence, mode collapse and overfitting of the generator. Thus, they barely meet the requirements for the inverse design of the materials.

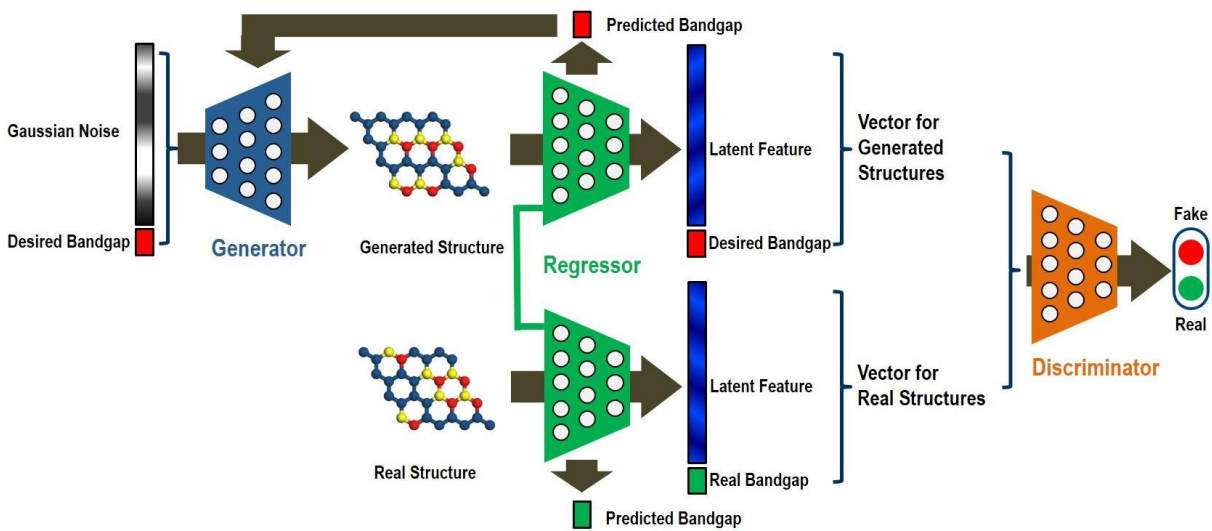

**Figure 1.** Architecture of the proposed regressional GAN (RGAN) for inverse structural design of doped graphene.

**Architectures of Regressor, Generator, and Discriminator**. To inversely design material structures, a regressional, semi-supervised GAN (RGAN) which can accurately associate the continuous labels (e.g. bandgaps of materials) with the generated material structures was proposed in this work (Fig. 1). In the RGAN, the generator is a 9-layer CNN (Fig. S1). Batch normalization[30] and leaky rectified linear unit (RELU) activation[31] were applied after each deconvolution step to increase the network stability and accelerate the training process. The generator synthesized graphene/BN structures from Gaussian noise (Z, a $128 \times 1$ vector) and appended desired bandgaps (Y). The architecture of the regressor as shown in Fig. S2 was modified from Google Inception V2.[32] It was trained by real material structures. After training, the regressor achieved a high prediction accuracy of > 98% when validated by the test datasets by following the procedure shown in our past work.[18] The generated structures and sampled real structures were then fed into the regressor for achieving two goals. The first is to accurately predict the bandgaps corresponding to given structures. The second is to output the latent features for both generated and real structures. The latent features were then concatenated to their corresponding bandgaps for forming two groups of vectors as the input datasets for training the discriminator. The discriminator is a simple network containing two fully connected layer intermediated by a dropout layer[33] to prevent overfitting (Fig. S3). It has two main functions. The first one is to maintain the function of automatic authentication, and the second one is to determine if a generated structure corresponds to a desired bandgap. The detailed description on these architectures as well as creation of training datasets is illustrated in Supplementary Note 1.

**Loss Functions of Regressor, Generator, and Discriminator**. The loss function of the regressor is defined as the $L_2$ loss between the bandgap of a real structures and the predicted bandgap output from the regressor.

$$Loss_R = L_2\left[Y, R(X)\right] \quad (1)$$

The $L_2$ loss function is defined as

$$L_2[X,Y] = \sum_i (x_i - y_i)^2 \quad (2)$$

The loss function for the generator includes two terms. The first one is the same as the least square GAN[34] (LSGAN). The second one is a regularization term from the regressor. The combined loss function is shown in the following.

$$Loss_G = \frac{1}{2} E_{z \sim P_z(z)} \left[ D\left(\pi\left(G(z,Y)\right), Y\right) - 1 \right]^2 + \lambda L_2\left(Y, R\left(G(z,Y)\right)\right) \quad (3)$$

In the first term of the Eq. 3, $E$ is the expectation function. $\pi(\cdot)$ is the latent feature generated from the last layer of the regressor upon given material structures, either from X or from the generator, G(z, Y). X, Y, respectively, is the structures and bandgaps of the real structures used for training. z is the random noise vector. D is the discrimination function. When feeding the regressor with the generated structures, the $L_2$ loss which indicates the difference between the predicted bandgaps and desired bandgaps was calculated, and then used as the regularization term in the loss function of the generator. The regularization term is reweighted using a weighting parameter λ and added to the loss function of the generator as shown in Eq. 3. λ was chosen as 25 in this work.

The loss function of discriminator is the same to the one used for LSGAN as shown in the following equation.[34]

$$Loss_D = \frac{1}{2} E_{X \sim P_{data}(X)} \left[ D(\pi(X), Y) - 1 \right]^2 \\ + \frac{1}{2} E_{z \sim P_z(z)} \left[ D(\pi(G(z, Y)), Y) \right]^2 \quad (4)$$

The subscript of $X \sim p_{data}(X)$ indicates that the structure was sampled from the real graphene/BN structures used for training, while the subscript of $z \sim p_z(z)$ indicates that the graphene/BN structure was synthesized by the generator.

**Results and Discussion**

We trained the RGAN with the datasets established by the DFT calculations as described in our previous work.[18] Representative 4 × 4 and 5 × 5 supercell systems of the graphene/BN structures and their corresponding material descriptors are shown in Fig. S4. These training datasets have 15,870 structures for 4 × 4 systems, and 80,647 structures for 5 × 5 systems, which represent 24.2% and 0.24% of all possible structures based on these systems, respectively. We traced the losses of the generator, regressor, and discriminator for the 4 × 4 and 5 × 5 systems during the training, and found that they gradually converged as the loss of the R was decreased. After 200 epochs of training, the R loss reached a steady state plateau (Fig. S5). The magnitude of loss for the regressor reaches as low as $10^{-3}$, indicating that the predicted bandgaps are quite close to the real bandgaps. The losses of the generator and discriminator also converged to stabilized values, indicating that the balance of the two networks has been reached.

Then the performance of RGAN was evaluated by studying the diversity and accuracy of the generated structures. Some examples of the generated structures with the varied desired bandgaps are shown in **Fig. 2**. The generator outputs some structures which have rotational or translational equivalence to the structures as shown in the training datasets. This situation occupies about 12% of all the generated structures for the $4 \times 4$ systems, but only $< 1\%$ for the $5 \times 5$ systems, which indicates that the RGAN was successfully trained without showing the problem of mode collapse.

All the generated $4 \times 4$ doped graphene structures shown in Fig. 2a-c are distinguishable from the real structures shown in the training datasets. These generated structures have bandgaps that are very close to the desired bandgaps as validated by the DFT validations. The structures are found to be diversified as indicated in Fig. 2d. It shows that the BN concentrations in the generated structures are ranging widely when given a single desired bandgap. It also confirms that the graphene/BN structures with the same BN concentration but different dopant topographies can have different bandgaps. For example, the graphene/BN structures with 50% BN have the bandgaps ranging from 0.9 eV to 1.7 eV. This also reflects the challenge in inverse structural design of the graphene/BN by the traditional ways due to such a large variation in the chemical spaces and properties.

The performance of the RGAN in generating $5 \times 5$ graphene/BN structures was also examined. Fig. 2e-g and Fig. S6 show examples of the generated structures whose bandgaps validated by the DFT calculations are also quite close to the desired bandgaps. **Fig. 2h** shows the diversity of the generated structures and the accuracy of their DFT bandgaps compared to the given desired bandgaps. The generation of these new structures with a reasonable accuracy indicates that the trained RGAN has successfully captured the distribution in the real structures used for training. We also successfully circumvented the barriers such as the mode collapse that often occurs during the GAN training, which can be attributed to the tactic of using the latent features as the input for training the discriminator.

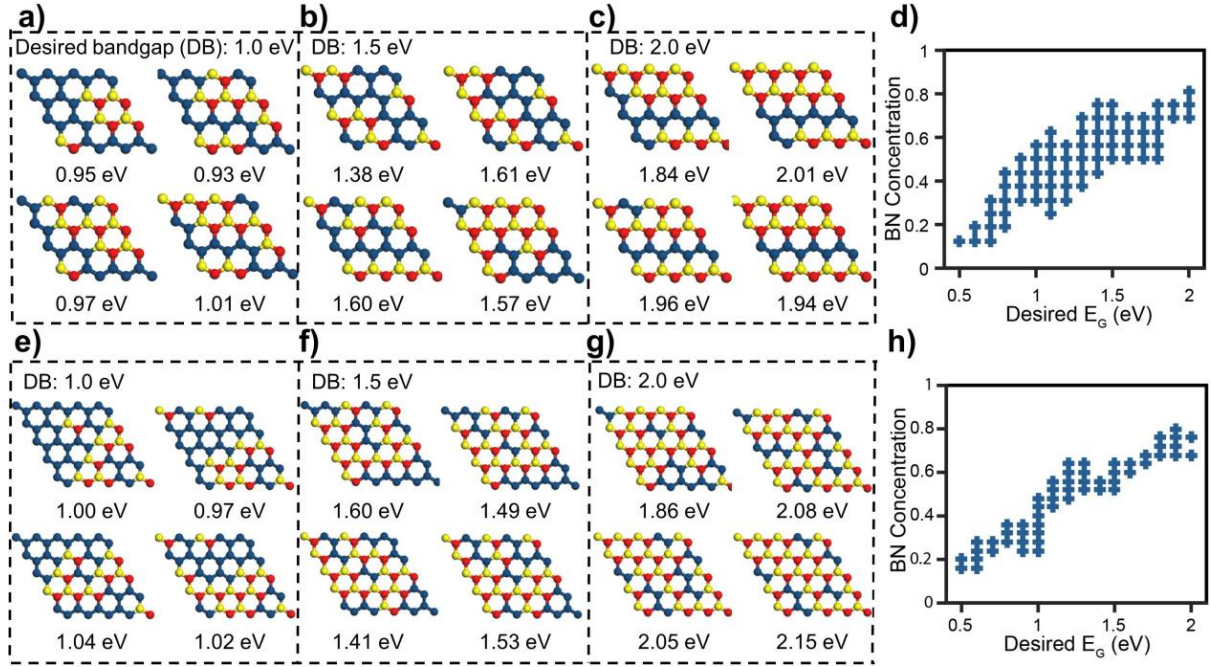

**Figure 2.** Examples of generated structures with desired bandgaps of **(a)** and **(e)** 1.0 eV, **(b)** and **(f)** 1.5 eV, **(c)** and **(g)** 2.0 eV for 4 × 4 supercell systems and 5 × 5 supercell systems, respectively. Distribution of boron-nitrogen (BN) concentrations in the generated structures for the 4 × 4 supercell systems **(d)** and 5 × 5 supercell systems **(h)**. The C, B, N atoms are colored with blue, red and yellow, respectively.

The performance of the RGAN was further analyzed by evaluating the bandgaps of the generated structures with given desired bandgaps (**Fig. 3**). Fig. 3a shows that the bandgaps of the generated structures have good linear relationship with the desired bandgaps. The correlation factor, $R^2$, is 0.82. We also studied the relative error between the bandgaps ($E_{GB}$) of the generated structures to the desired bandgaps ($E_{DB}$), which is calculated as $|E_{DB}-E_{GB}|/E_{DB}$. Fig. 3b shows that 55% of the generated structures have bandgaps within a 10% relative error. The number increases to > 80% if within a 20% relative error. The fractional mean absolute error ($MAE_F$) (Supplementary Note 2) between the generated bandgaps and the desired bandgaps is as low as 9.67% for the $4 \times 4$ systems. For the $5 \times 5$ systems, Fig. 3c shows that $R^2$ reaches 0.87, which is as good as the supervised prediction task by the CNNs.[18] Fig. 3d shows that ~ 50% of the generated structures whose bandgaps fall within a relative error of 10% of the desired bandgaps, while < 20% of the structures have a relative error of 20% or more. The $MAE_F$ of the $5 \times 5$ systems is 11.6%. Although it is a little higher than that of the $4 \times 4$ systems, the performance of RGAN is still well maintained as the size of the systems increases. As the increased system size is generally accompanied by the performance degradation,[18] these results suggest the potentials of the RGAN in exploring much larger material spaces with an even smaller sampling ratio, for example, a 0.24% sampling ratio for the $5 \times 5$ systems.

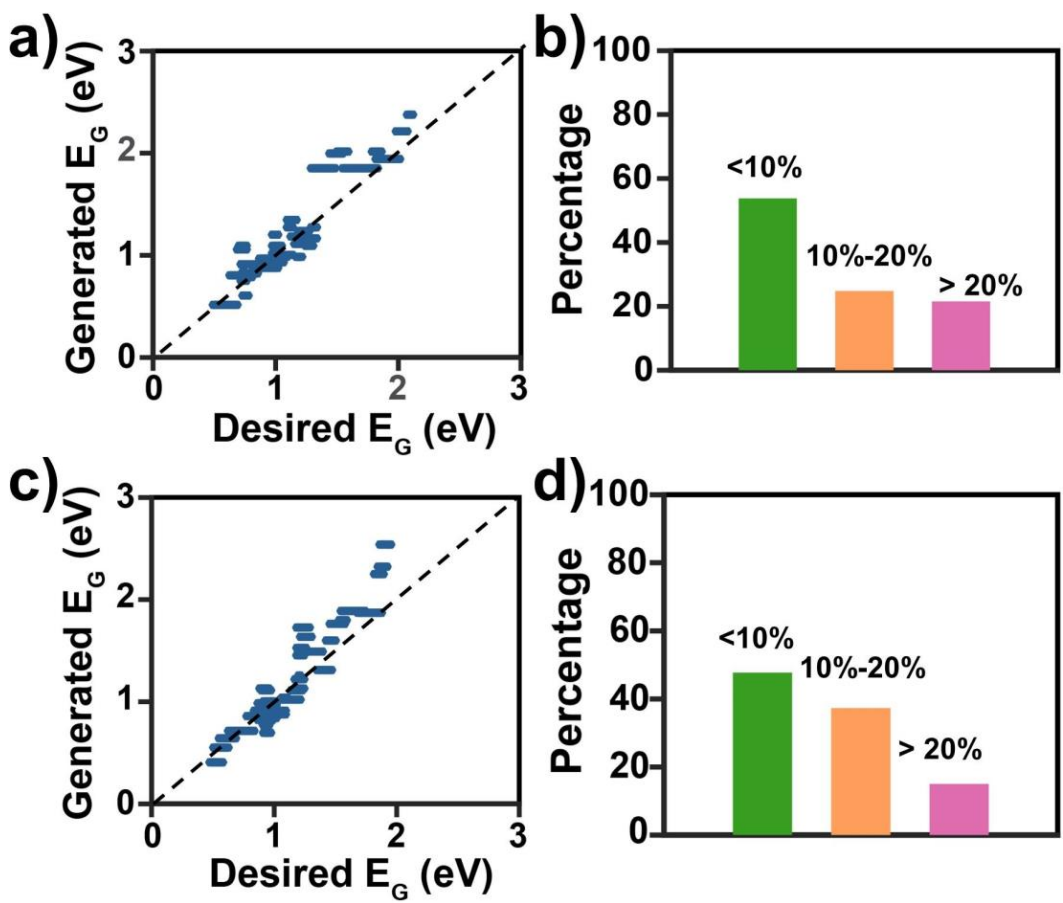

**Figure 3.** (a) Bandgaps of generated structures by DFT vs. desired bandgaps for $4 \times 4$ supercell systems, $R^2$=0.82. (b) Distribution of generated bandgaps for $4 \times 4$ supercell systems, $MAE_F$=9.67%. (c) Bandgaps of generated structures by DFT vs. desired bandgaps for $5 \times 5$ supercell systems, $R^2$=0.87. (d) Distribution of generated bandgaps for $5 \times 5$ supercell systems, $MAE_F$=11.6%.

The successful generation of the graphene/BN structures with desired bandgaps by the proposed RGAN motivates us to hypothesize that the latent features encoded by the regressor assists to associate the generated structures with given continuous quantitative bandgaps. To test this hypothesis, a key idea is to validate whether the latent features of the structures synthesized from the generator has successfully followed the statistical distribution of the real structures used for training. Thus, we performed dimension reduction analysis on both of them through the principal component analysis (PCA)[35] and modified locally linear embedding (MLLE)[36] (Supplementary Note 3). Fig. 4a shows the mapping of first and second components of PCA ($PCA_X$ and $PCA_Y$) of the latent features from 1000 4 ×4 real structures and 1000 4 ×4 generated structures. They are highly overlapped, suggesting that the RGAN has successfully learned the distribution of the latent features from the real structures when synthesizing new structures. Fig. 4b shows that the $PCA_X$ has a good linear relationship to the bandgaps (labels). We hypothesize that $PCA_x$ is the term that is strongly correlated with the BN concentration. Such a relationship indicates that the generator may catch the physical law that the bandgap increases with the increase of the BN concentration, which agrees well with the results suggested from the real structures used for training (Fig. 4c). The curve of the bandgap vs. $PCA_Y$ shows that the generated structures can be grouped to two regions according to the bandgaps (Fig. 4d). The structures with high bandgaps (> ~1.0 eV) are in one region, while the ones with low bandgaps (< ~1.0 eV) are in the other.

In addition to the PCA, we also performed the MLLE analysis. The two sets of data points from the real and generated structures form an almost perfectly overlapped parabolic curve (Fig. 4e). It suggests that the MLLE can well capture the main pattern of the relationship between structural configurations and bandgaps. The mapped bandgaps of the generated structures vs. the first and second components of the MLLE results are shown in (**Fig. S7a-b**). They illustrate that these data points have similar trend as the ones shown in the PCA. It also shows that the bandgaps monotonically increase as the $MLLE_x$ increases. Further analysis on the structural evolution along the path shown in the MLLE curve suggests that the location and concentration of BN pairs gradually change as the bandgap values increases (Fig. 4f). Their neighboring structures share similar BN topographies with limited discrepancy.

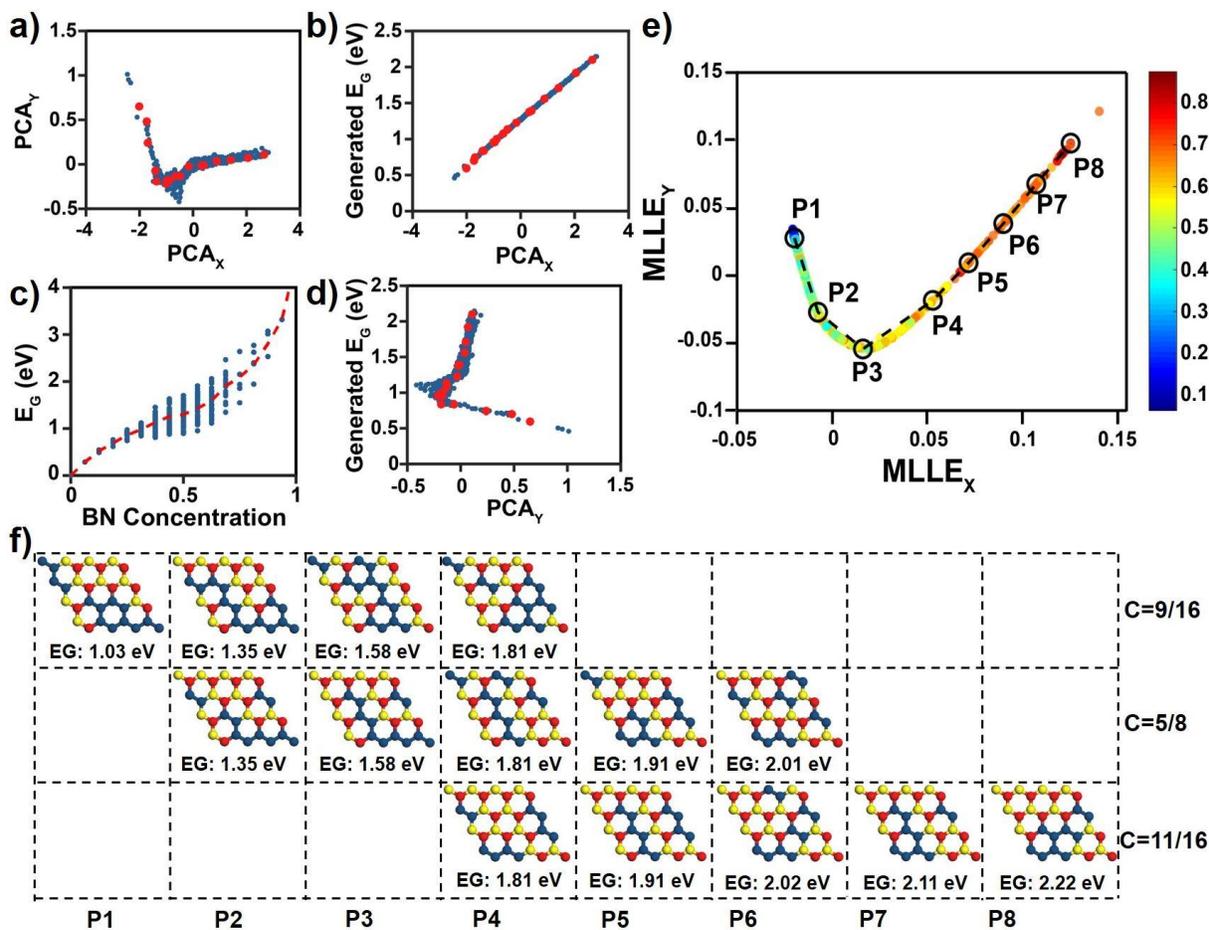

**Figure 4.** PCA and MLLE analysis on the latent features of 4 × 4 supercell systems encoded by the regressor. **(a)** Mapping of $PCA_X$ and $PCA_Y$ components. The Blue dots are the training data, and the Red dots are the generated data. **(b)** Relationship of generated bandgap vs. $PCA_X$. **(c)** Bandgaps of real structures vs. BN concentrations. **(d)** Relationship of generated bandgap vs. $PCA_Y$. **(e)** Mapping of $MLLE_X$ and $MLLE_Y$ from real and generated structures. The dots represent points in the datasets. The colors of dots represent the BN concentration in graphene/BN structures. **(f)** Structural evolution along the points (P1-P8) marked in **(e)**. C indicates the BN concentration.

**Conclusion**

In this work, we demonstrated a novel RGAN architecture for inverse structural design of a representative 2D material upon the continuous quantitative property label. The extraction and usage of the latent features encoded from the regressor as the input for the discriminator was critical to successfully train the RGAN. The generated structures have a similar distribution with the real structures used for training. Statistically, the bandgaps of the generated structures are close to the desired bandgaps, exhibiting $MAE_F$ of ~ 9.67% and 11.6% for 4 × 4 and 5 × 5 systems, respectively. Through this illustrative example, we verified the power of the proposed RGAN in accurately designing materials with on-demand properties in an autonomous manner. Considering that the graphene is still fascinating, doping or alloying with heterogeneous atoms with the atomic level precision would open a new possibility. This work could afford a platform for highly expected autonomous material design enabled by big material data and artificial intelligence.

**Acknowledgement**

J. L. and M.X. acknowledge financial support from US Department of Energy (Award number: DE-FE0031645) with Program Manager Karol K. Schrems, National Science Foundation (Award number: 1825352) with Program Manager Khershed P. Cooper. J. C. acknowledges National Science Foundation (Award numbers: DBI1759934 and IIS1763246). The DFT calculations were performed on the HPC resources at the University of Missouri Bioinformatics Consortium (UMBC), supported in part by NSF (award number: 1429294).

**Author contributions**

Y. D. designed material descriptive system, performed the DFT calculations, and did the data analysis. D. L. designed and trained the RGAN as well as performing the PCA. C. Z. assisted Y.

D. in DFT calculations and performed data analysis. H. W. performed the MLLE analysis. M. X. contributed to the discussion of the project. J. C. initially proposed the idea of applying GAN to graphene design and supervised C. W. to explore the idea by testing ACGAN. J. L. organized the scopes of the project and oversaw all phases of the project. Y. D., J. L., and D. L. wrote the manuscript. All the authors commented on the manuscript.

**Competing interests**

The authors declare no competing interests.

**Materials & Correspondence:** Correspondence and material requests should be addressed to J. L. (LinJian@missouri.edu).

**Data availability:** All data needed to evaluate the conclusions in the paper are present in the paper and/or the Supplementary Materials. Additional data related to this paper may be requested from the authors. Supplementary information is available at XXXX website.

[1]Department of Mechanical and Aerospace Engineering,
[2]Department of Electrical Engineering and Computer Science,
[3]Department of Physics and Astronomy,
University of Missouri, Columbia, Missouri 65211, USA
*E-mail: LinJian@missouri.edu (J. L.)
#Authors contributed equally to this work.

**Supplementary Notes**

**Supplementary Note 1. Design and Training of Regressional GAN (RGAN)**

**Architecture of the Generator**. The architecture of the generator is shown in **Fig. S1**. To generate a graphene/BN structure with a desired bandgap, the bandgap was appended to a randomly generated noise vector z with a dimension of 128×1 and then fed into the generator. The first five modules of the network consist of an inverse convolutional layer with 512 3 × 3 filters which slide the reshaped input with a stride of 1, a batch normalization layer. The activation function is a leaky rectified linear unit (RELU) with the parameter of 0.2. Two modules subsequently follow the above five modules with reduced number of filters from 512 to 256 and then to 128. A different number of strides (either 1×1 or 2×2) were used to ensure that the dimension of the generated structure matches the one represented for the real structure. There is a final inverse convolutional layer to ensure that the only one structure is generated at each cycle. The *tanh* activation function for the final layer outputs continuous numbers ranging from -1 to 1. Before sending the outputs into the regressor, the numbers less than 0 were replaced with

-1 to indicate a C-C pair, while the numbers larger than 0 were replaced by 1 to represent a BN pair when forming matrices representing the generated graphene/BN structures.

**Architecture of the Regressor**. The structure of the regressor as shown in **Fig. S2** was modified from Google Inception V2.[1] The inception modules have been used in the network with each module activated by RELU. Some of the modules are followed by an additional max-pooling layer. The output from the module is flattened and enters a RELU activated dense layer with 512 nodes. Eventually, after an extra dense layer of 32 nodes, the final output layer with 1 node can output predicted bandgaps. Also, we extracted the intermediate output, $\pi$, from the fully connected layer (L) before the output layer. The outputs are the latent features that represent the distribution of the generated and real graphene/BN structures.

**Architecture of the Discriminator**. The discriminator has two functions (**Fig. S3**). The first one is to maintain the function of automatic authentication, and the second one is to determine if a generated structure corresponds to a desired bandgap. In order to append the information of the bandgaps, the desired form of the data fed into the discriminator should be a single vector rather than a high-dimension tensor. Therefore, instead of directly feeding a real or a generated structure to the discriminator, a combined vector by directly concatenating the latent feature ($\pi$) of the structure and its corresponding bandgap was employed as the input to the discriminator. The discriminator is trained to tell whether the concatenated vector is from a real structure or a generated structure. In order to balance the whole architecture of RGAN, the discriminator that has only one intermediate dense layer with 256 nodes, followed by a leaky RELU activation layer, was built. The output layer is a single-node dense layer activated by sigmoid function, which forces the output ranging from 0 to 1, indicating fake or real, respectively. By this way, the two essential functions of the discriminator mentioned above are preserved.

**Training Datasets for RGAN**. We performed the *ab initio* DFT calculations by the QUANTUM ESPRESSO package.[2] The ultra-soft projector-augmented wave (PAW) pseudopotential[3] was used to characterize the valence electrons-ions interaction and the Perdew-Burke-Ernzenhof (PBE) schema[4] was used to describe the exchange-correlation energy. The density functional plane wave was cut off at energy of 400 eV. The Brillouin zone was meshed by 8 × 8 × 1 in the k-point space using the Monkhorst-Pack scheme.[5] To perform high throughput calculation, 4 × 4 and 5 × 5 matrices for representing the graphene/BN structures were randomly generated. The elements in the matrices were randomly selected between "-1" and "1", where "-1" stands for the carbon-carbon (C-C) pairs and "1" stands for the boron-nitrogen (BN) pairs. After the matrices were generated, the DFT calculations afford the bandgaps of these structures. The datasets containing the structures and corresponding bandgaps were created for subsequent training of RGAN. These training datasets have 15870 structures for 4 × 4 systems, and 80647 structures for 5 × 5 systems, which represent 24.2% and 0.24% of all possible structures from their corresponding systems, respectively.

**Training of RGAN**. The RGAN was designed and trained by Google's Tensorflow API version 1.14.0. Adam, with a learning rate of $1e^{-4}$, was selected as the optimizer for the generator, discriminator, and regressor. A $\beta_1$ regularization term with a value of 0.5 was added to both the optimizers for the generator and discriminator. The RGAN was then trained with 256 epochs and a batch size of 105 samples in each cycle.

**Supplementary Note 2. Evaluation Metrics**

The discrepancy between the DFT calculated bandgaps of the generated structures and the desired bandgaps were evaluated by standard statistical metrics including the explained variance ($R^2$), mean absolute error (MAE), and root mean squared error (RMSE), metrics are provided as follows.

$$R = \frac{\sum_{i=1}^{N}(X_i - \bar{X})(Y_i - \bar{Y})}{\sqrt{\sum_{i=1}^{N}(X_i - \bar{X})^2}\sqrt{\sum_{i=1}^{N}(Y_i - \bar{Y})^2}} \quad (S1)$$

$$\text{MAE} = \frac{1}{N}|X_i - Y_i| \quad (S2)$$

$$\text{RMSE} = \sqrt{\frac{1}{N}\sum_{i=1}^{N}(X_i - Y_i)^2} \quad (S3)$$

where X is the desired bandgap and Y is the DFT calculated bandgap of the generated structure. $\bar{X}$ is the mean value of X, and analogously for $\bar{Y}$. N is the total number of the structures. MAE and RMSE can be further normalized as follows after being calculated.

$$\text{MAE}_F = \frac{1}{N}\left|\frac{X_i - Y_i}{X_i}\right| \quad (S4)$$

$$\text{RMSE}_F = \sqrt{\frac{1}{N}\sum_{i=1}^{N}\left(\frac{X_i - Y_i}{X_i}\right)^2} \quad (S5)$$

**Supplementary Note 3. Principal component analysis (PCA) and Modified Locally Linear Embedding (MLLE)**

The principal component analysis (PCA) performs an orthogonal linear transformation on data to pursue the greatest variance on its principle components. It is commonly used in many machine learning studies for dimension reduction, variable interpretation, and data visualization.[6-8] It was worth pointing out that PCA efficacy would be degraded if the internal data patterns are too nonlinear or/and strongly correlated.[9] Locally linear embedding (LLE) preserves the neighboring relationship of the data points. LLE uses the locally linear fit method to rebuild nonlinear structure, which is different from the PCA. The LLE offers more power in revealing hidden features of the high-dimensional data.[10] For a given data vector X, the LLE defines the cost of reconstruction as:

$$\varepsilon(W) = \sum_i \left| X_i - \sum_j W_{ij} X_j \right|^2 \quad (S6)$$

where W is the weighting vectors. The embedding cost function is:

$$\Phi(Y) = \sum_i \left| Y_i - \sum_j W_{ij} Y_j \right|^2 \quad (S7)$$

where Y is the principle vectors in the reduced dimension. The modified locally linear Embedding (MLLE) was later proposed to improve the robustness of LLE.[11] It adopts the modified cost function as follows:

$$\Phi(Y) = \sum_i \sum_k \left| Y_i - \sum_j W_{ij}^{(k)} Y_j \right|^2 \quad (S8)$$

where $W^{(k)}$ are a set of linearly independent weighting vectors. LLE and MLLE were shown to perform well in nonlinear dimension reduction problems, with a wide variety of applications in facial/image recognition and word clustering.

**Supplementary Figures**

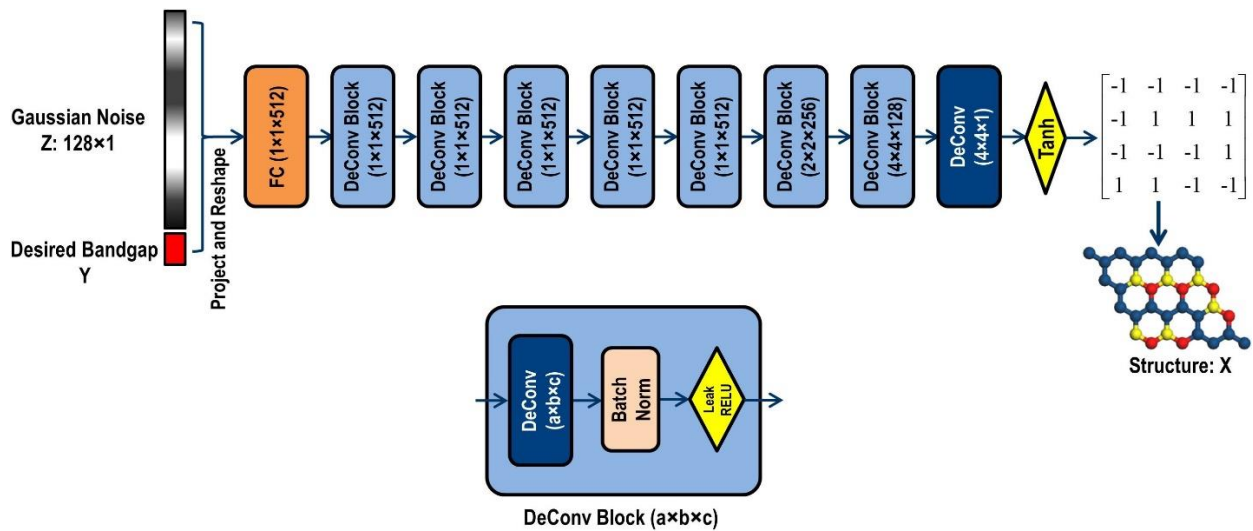

**Figure S1.** Architecture of the generator. Bottom panel is the architecture of the de-convolutional block for the generator. The generator for 5×5 structure has no padding on the last layer to ensure that the output after inverse convolutional layers is $5 \times 5 \times 1$.

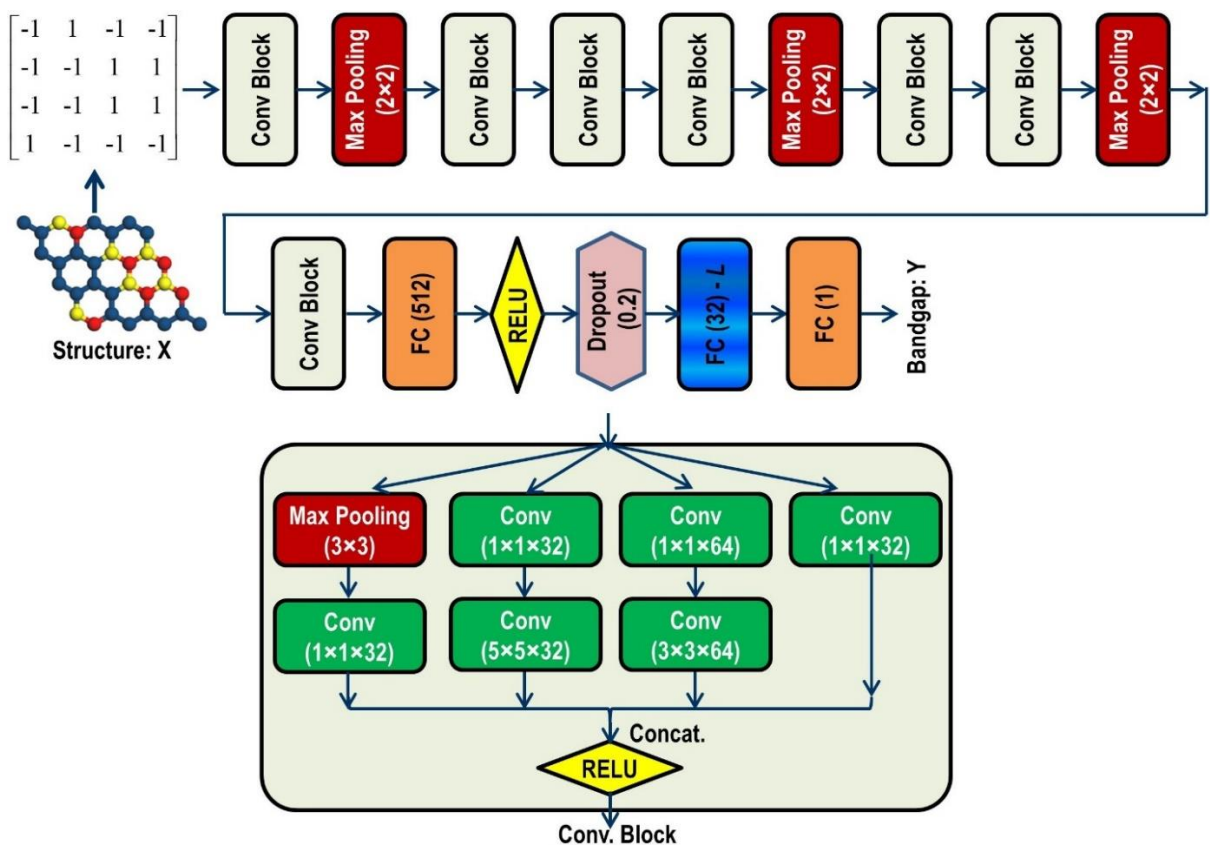

**Figure S2.** Architecture of the regressor. The *L* layer before the output layer is colored with blue. It outputs the latent features of the generated and real graphene/BN structures, which are fed into the discriminator. Bottom panel schematically show the architecture of the convolutional block for the regressor.

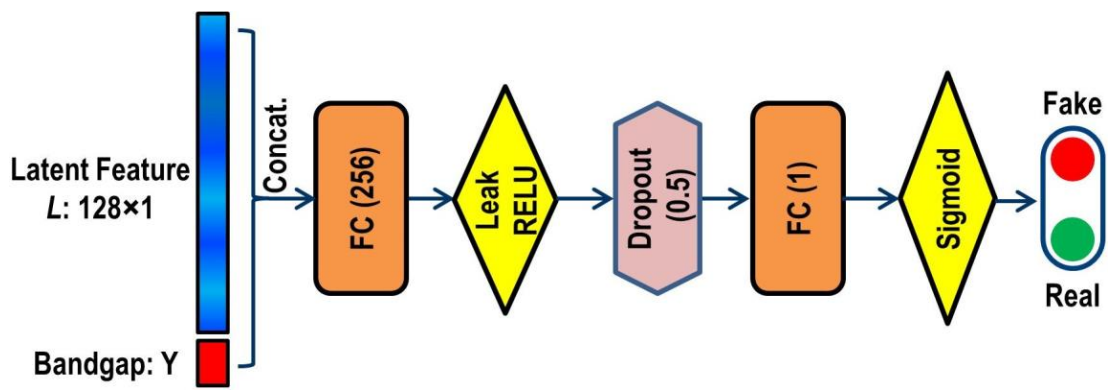

**Figure S3.** Architecture of the discriminator.

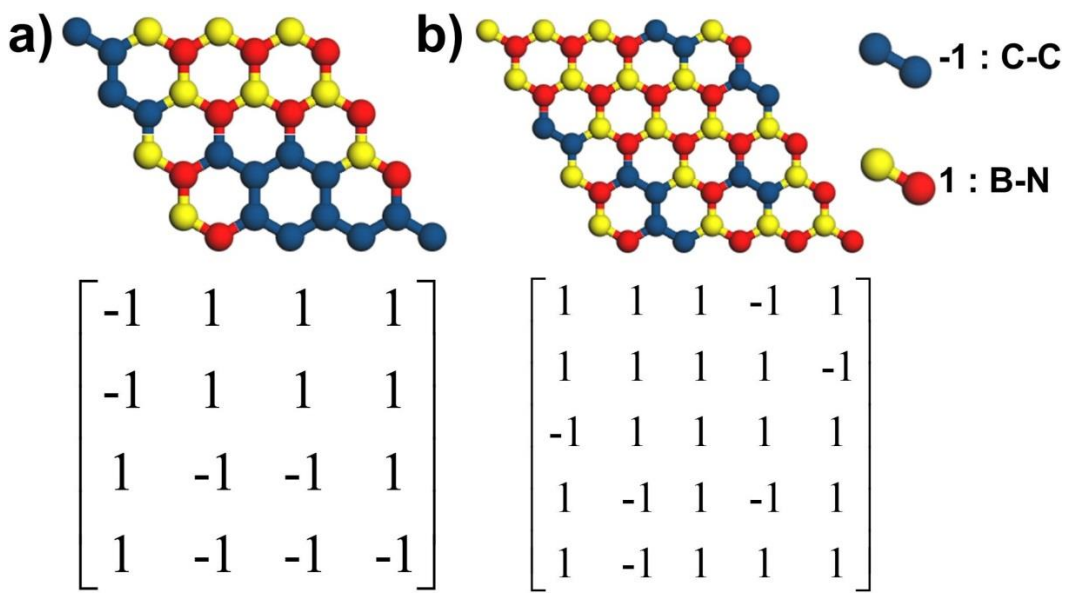

**Figure S4.** Structures of **(a)** 4×4 **(b)** 5×5 graphene/BN hybrid systems and their corresponding representation matrices for the descriptors. The element "-1" represents a carbon-carbon pair and the element "1" represents a boron-nitride pair.

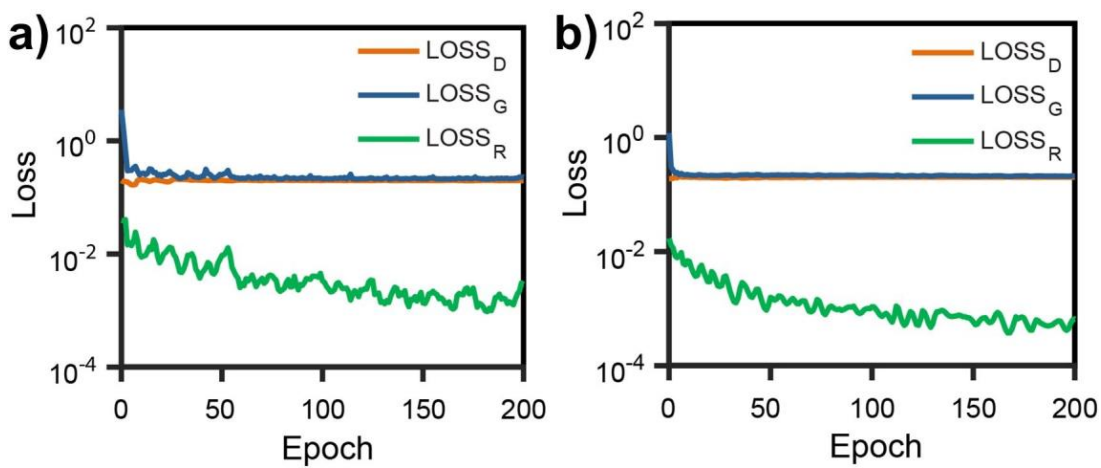

**Figure S5.** Loss evolution of discriminator (D), generator (G), and regressor (R) for **(a)** 4×4 supercell systems and **(b)** 5×5 supercell systems during the training processes.

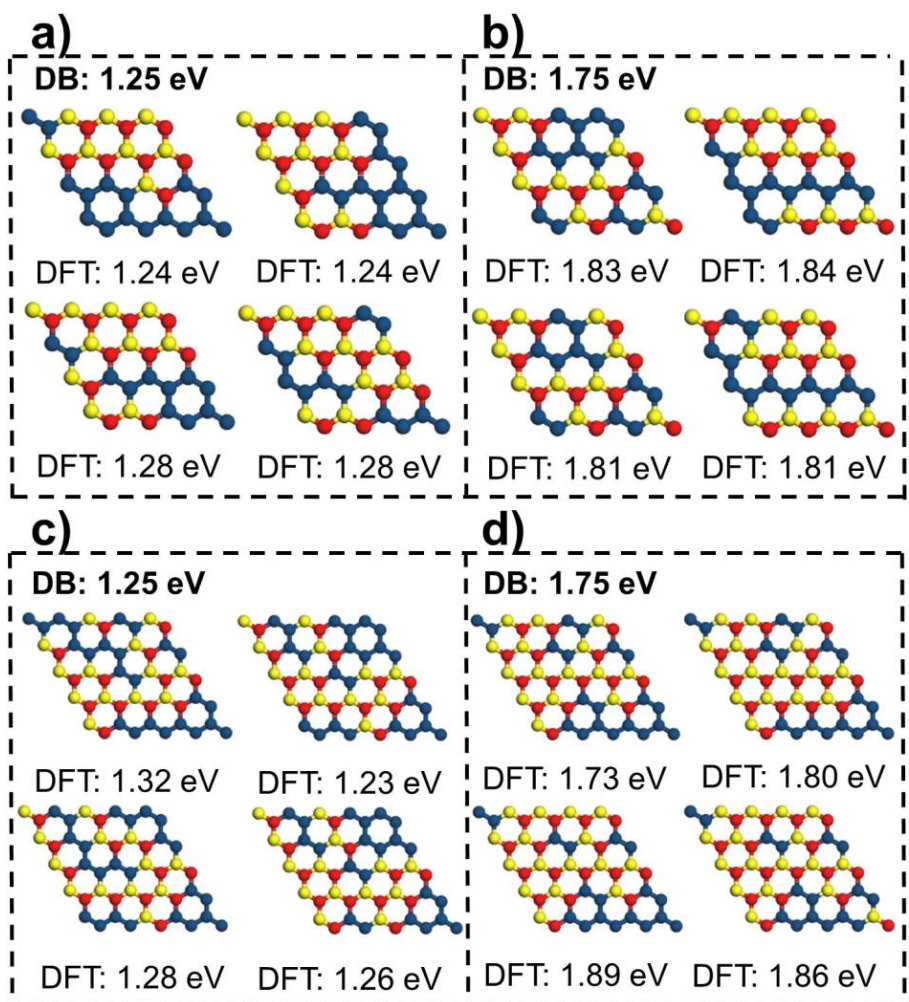

**Figure S6.** Typical generated structures with desired bandgaps **(a)** 1.25 eV, **(b)** 1.75 eV for $4 \times 4$ supercell systems. Typical generated structures with desired bandgaps **(c)** 1.25 eV, **(d)** 1.75 eV for $5 \times 5$ supercell systems.

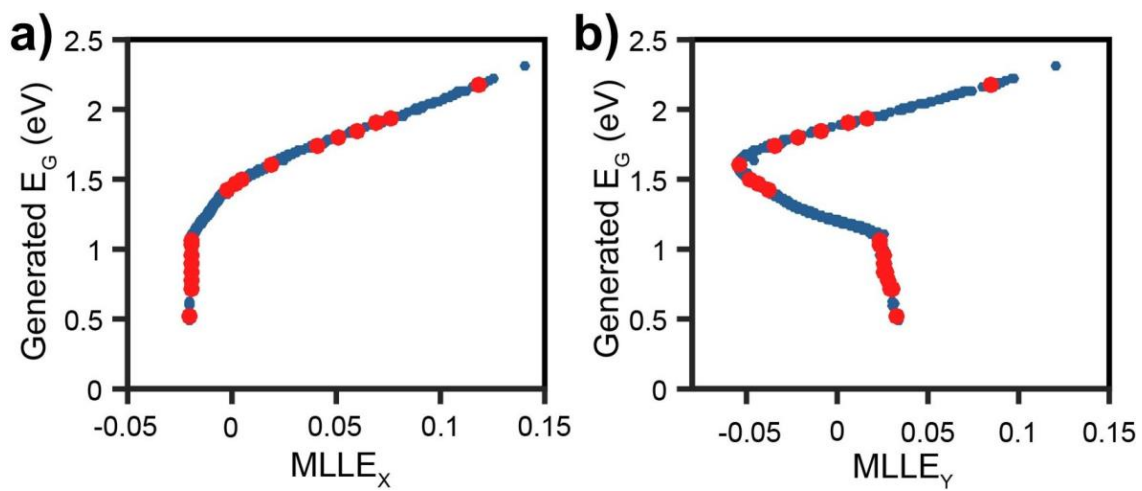

**Figure S7.** MLLE results of the latent features from the 4 × 4 supercell systems. Generated bandgap vs. **(a)** first and **(b)** second principle components. The Blue dots are the data from the training structures, and the Red dots are data points from the generated structures.

**Supplementary References**